\documentclass[prl,floats,showpacs,aps,twocolumn,floatfix,amsmath,reprint]{revtex4-1}
\usepackage{graphicx,amsbsy}
\begin{document}
\title{Parity anomaly and Landau-level lasing in strained photonic honeycomb lattices}

\author{Henning Schomerus}
\author{Nicole Yunger Halpern}
\affiliation{Department of Physics, Lancaster University, Lancaster,  LA1 4YB, United Kingdom}

\date{\today}

\begin{abstract}
We describe the formation of highly degenerate, Landau-level-like
amplified states in a strained photonic honeycomb lattice in which
amplification breaks the sublattice symmetry. As a consequence of the
parity anomaly, the zeroth Landau level is localized on a single
sublattice and  possesses an enhanced or reduced amplification rate.
The spectral properties of the higher Landau
levels are constrained by a generalized time-reversal symmetry. In the
setting of two-dimensional photonic crystal lasers, the anomaly directly affects
the mode selection and lasing threshold while in three-dimensional
photonic lattices it can be probed via beam dynamics.
\end{abstract}
\maketitle

Nonuniform deformations of the honeycomb lattice of graphene result in a
pseudomagnetic field which deflects particles in analogy to the Lorentz
force, with small amounts of strain producing fields that are large
enough to create well-defined Landau levels in the low-energy range of
the spectrum, in absence of any physical magnetic field
\cite{guinea,vozmediano,levy,yan}. Here we describe how the addition of
gain in an analogous photonic setting results in the formation of
highly degenerate amplifying Landau levels, which can provide the
platform for a laser with macroscopic mode competition. The spectral
properties of these levels become intriguing when the gain breaks the
sublattice symmetry. Due to the parity anomaly
\cite{jackiw,semenoff,haldane}, the amplification of the zeroth Landau
level is dictated by one of the two sublattices, which here is selected
depending on the strain orientation. Moreover, a reflection symmetry
enforces that the instances of this level in the two $\mathbf{k}$-space
valleys behave identically.
In contrast, the higher Landau levels are constrained by a generalized
time-reversal symmetry. Their amplification rate equals the average rate
on the two sublattices, up to a finite threshold of the imbalance at
which two levels coalesce and their rates bifurcate.

These observations allow to detect the parity anomaly via the anomalous
amplification or decay of the zeroth Landau level. When the system is
operated as a two-dimensional photonic crystal laser, the lasing
threshold is set either by the zeroth or by the first Landau level, with
the selection dictated by the strain orientation and signature of the
amplification imbalance. We also describe how the anomalous behavior of
the zeroth Landau level can be probed via the beam dynamics in a
three-dimensional photonic lattice.

\begin{figure}
\includegraphics[width=\columnwidth]{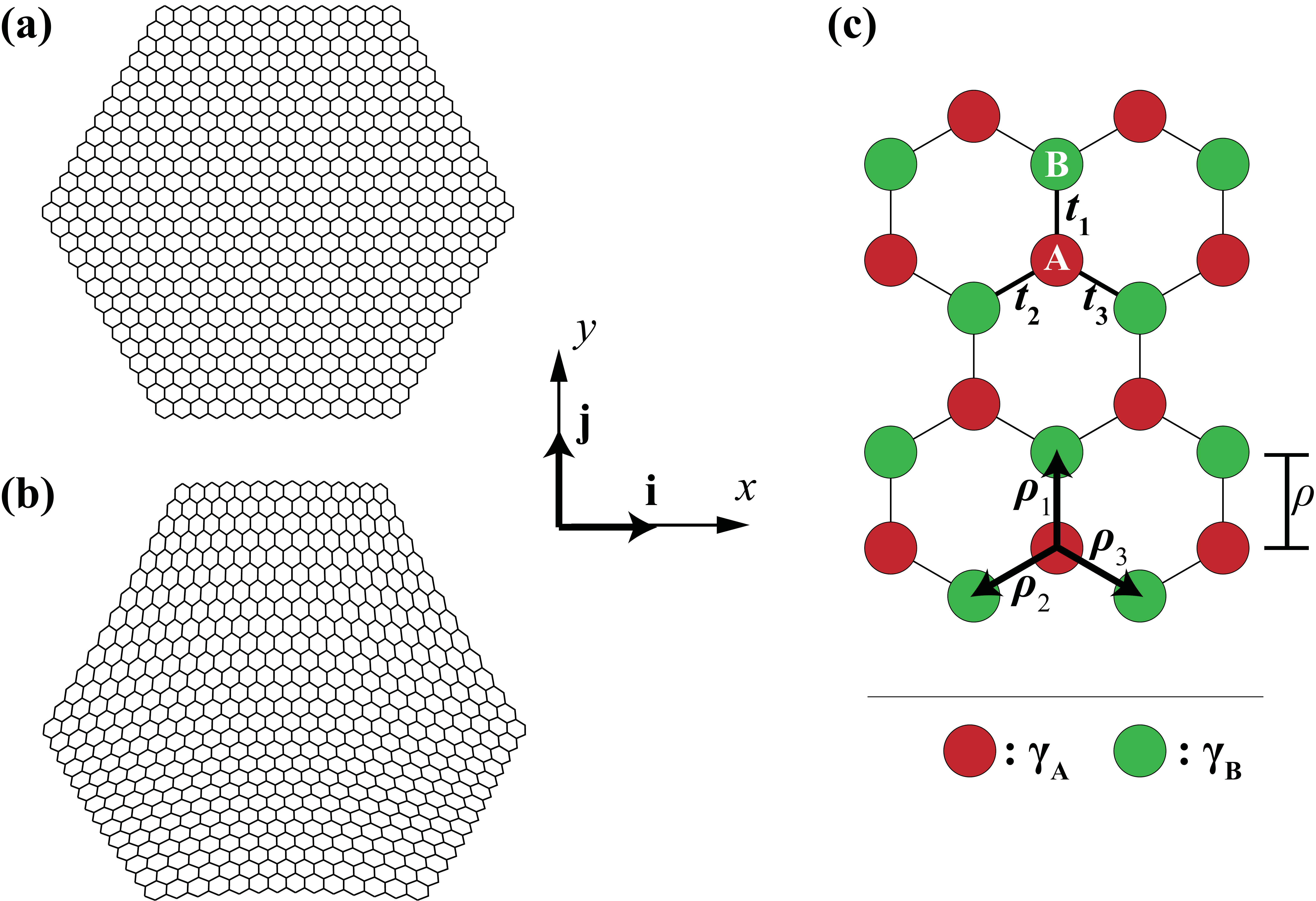}
\caption{\label{fig1} (Color online)
(a) Segment of a honeycomb lattice,  with vertices  representing states in a two-dimensional photonic crystal or weakly coupled optical fibers in a three-dimensional setting.
(b) Sketch of a deformed arrangement which results in a constant pseudomagnetic field.
(c) We investigate the interplay of this field with amplification and absorption that breaks the sublattice symmetry. The two sublattices A and B have amplification rates $\gamma_A$ and $\gamma_B$, respectively (negative values correspond to absorption).
The pseudomagnetic field resulting from the strain is modeled via smooth coupling functions $t_l$ whose definition \eqref{eq:tfunc} involves the bond vectors $\boldsymbol{\rho}_l$, $l=1,2,3$.}
\end{figure}

\begin{figure}
\includegraphics[width=\columnwidth]{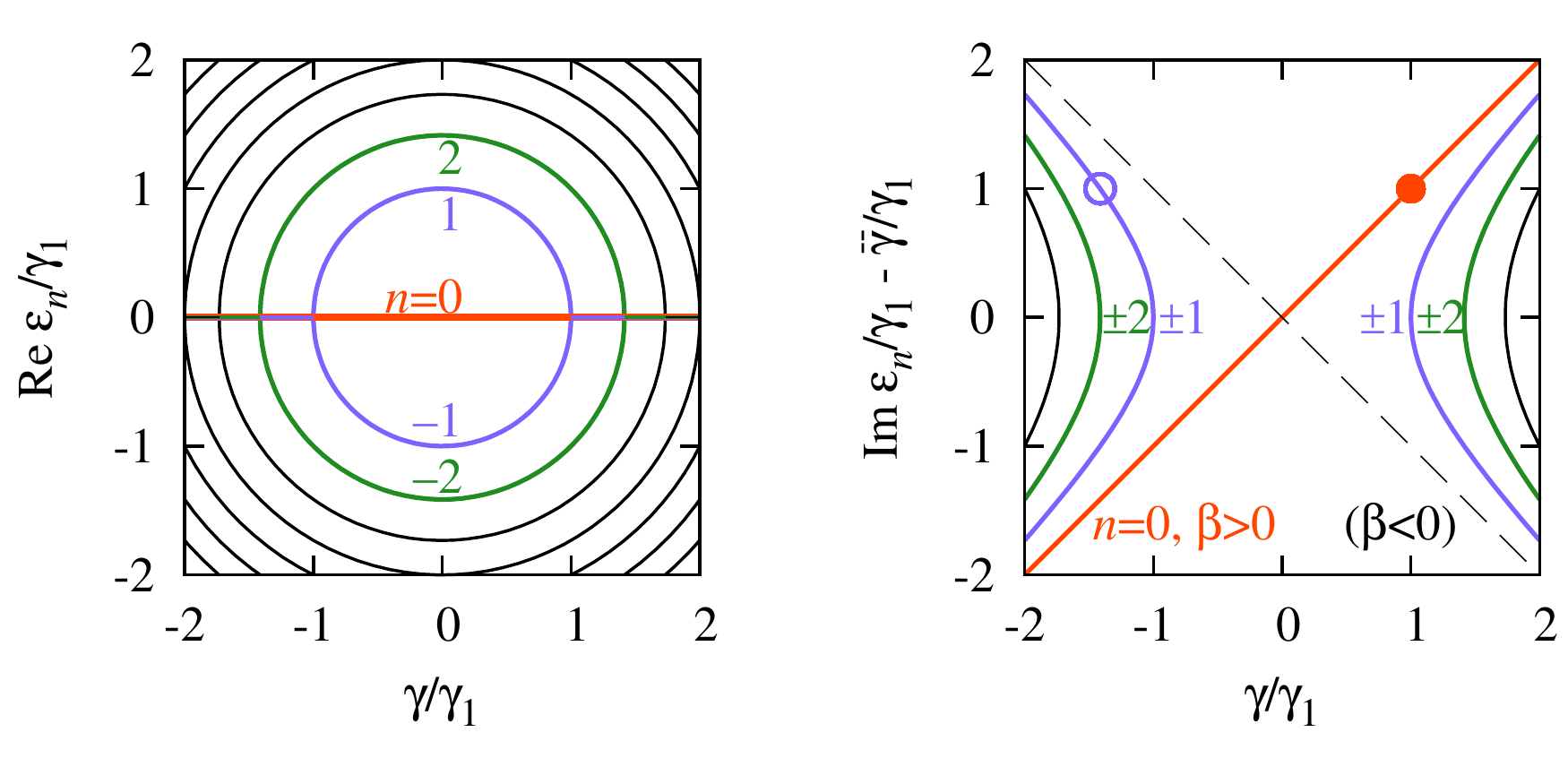}
\caption{\label{fig2} (Color online)
Dependence of the Landau level spectrum \eqref{eq:zerothll}, \eqref{eq:higherll} on the amplification imbalance $\gamma=(\gamma_A-\gamma_B)/2$, for strain leading to a pseudomagnetic field of strength $\beta$. (a) Real part, which vanishes for the zeroth Landau level, as well as for the other Landau levels beyond their bifurcation thresholds $\gamma_n$, Eq.~\eqref{eq:higherllgamma}. (b)
Imaginary part, which becomes finite beyond the bifurcation. Because of the parity anomaly, the zeroth Landau level breaks the symmetry of the spectrum as it is located on the A sublattice for $\beta>0$ (thick solid line) while it is located on the B sublattice for $\beta<0$ (thin dashed line). The circles indicate the lasing threshold when $|\gamma|$ is increased
at fixed average absorption $\bar\gamma=-\gamma_1$. For $\gamma>0$ this threshold is set by the zeroth Landau level, $\gamma_L=\gamma_1$ (solid circle), while for $\gamma<0$ it is set by the first Landau level, $\gamma_L=-\sqrt{2}\gamma_1$  (open circle).
}
\end{figure}

\emph{Model of a strained active photonic honeycomb lattice.---}We
specifically consider the photonic system sketched in Fig.~\ref{fig1}.
Panel (a) shows a segment of a honeycomb lattice, with vertices
representing weakly coupled optical fibers in a three-dimensional
photonic lattice \cite{peleg} or a set of basis states for a suitable
spectral range in a two-dimensional photonic crystal \cite{pc,raghu}. The
honeycomb lattice consists of two sublattices, A sites and B sites, which
we equip with different amplification or absorption rates. This is
motivated by recent works on optical realizations
\cite{makris,experiment1a,experiment1b,tachyons,ramezani,Regensburger} of
non-hermitian $\mathcal{PT}$-symmetric quantum mechanics \cite{bender}.
In the present setting, $\mathcal{P}$ stands for the inversion about the
center of a hexagon, which maps A sites to B sites and thus inverts the
amplification imbalance; $\mathcal{T}$ corresponds to complex conjugation
and converts amplification into absorption, which also inverts the
imbalance. Panel (b) sketches an inversion-symmetry-breaking deformed
arrangement which results in a constant pseudomagnetic field whose
interplay with the symmetry-breaking effects of amplification and
absorption we are interested in. Panel (c) illustrates the microscopic
modeling of these effects. The sublattices carry amplification rates
$\gamma_A=\bar\gamma+\gamma$ and $\gamma_B=\bar\gamma-\gamma$,
respectively, where $\bar\gamma$ is the average rate and $\gamma$
quantifies the imbalance. The rates $\gamma_{A}$ and $\gamma_B$ may be
negative, in which case they signify absorption. Strain results in a
spatial variation of the coupling terms $t_{ab}$ between neighboring A
and B sites, which we parameterize as $t_{ab}=t_{l}(\mathbf{r}_a)$, where
$\mathbf{r}_a$ is the unstrained position of the A site and $l=1,2,3$
indicates the orientation along the unstrained bond vectors
$\boldsymbol{\rho}_l$ ($|\boldsymbol{\rho}_l|=\rho$ is the unstrained
nearest-neighbor distance). The typical magnitude of the coupling terms
is denoted as $t_0$.

We focus on a spectral range where the unstrained passive lattice
displays a conical band structure \cite{raghu}. Based on the descriptions of strained graphene
\cite{guinea,vozmediano,wallace,Sasaki,saito} and  photonic honeycomb
lattices or crystals
\cite{peleg,pc,raghu,makris,experiment1a,experiment1b,tachyons,ramezani},
the resulting Lorentz force and the effects of amplification and
absorption are then captured by a Dirac equation with Hamiltonian
\cite{supmat}
\begin{equation}\label{eq:dirac}
\mathcal{H}=\left(\begin{array}{cc} i\gamma_A & v (\sigma P_x-iP_y)\\ v (\sigma P_x+iP_y) & i\gamma_B \end{array}\right),
\end{equation}
$v=3t_0\rho/2$, $P_x=-i\partial_x-A_x$, $P_y=-i\partial_y-A_y$, where
\begin{equation} \label{eq:a}
 \mathbf{A}=\sigma\frac{1}{3\rho
t_0}(2t_1-t_2-t_3)\mathbf{i}+\sigma\frac{1}{\sqrt{3}\rho
t_0}(t_2-t_3)\mathbf{j}
\end{equation}
is the pseudomagnetic vector potential. This Hamiltonian applies to a
continuous spinor wave function
$(\varphi_{A}(\mathbf{r}),\varphi_{B}(\mathbf{r}))^T$ which is obtained
by separating out rapid fluctuations with wave vector
$\mathbf{K}_\sigma=\sigma (4\pi/3\sqrt{3}\rho)\mathbf{i}$, $\sigma=\pm
1$, where $\sigma$ distinguishes two independent valleys; these valleys are
related by the $\mathcal{P}$ and $\mathcal{T}$ symmetries of the
unstrained passive system \cite{raghu}. The eigenvalues $\varepsilon$ of
${\cal H}$ determine the frequencies $\omega$ of quasibound states in the
two-dimensional setting \cite{pc,raghu} or the propagation constant
$ck_z$ along the third direction in the three-dimensional setting
\cite{peleg,makris}. Eigenvalues with a positive imaginary part
correspond to amplified states  (in time or along the propagation
direction), while those with a negative imaginary part correspond to
decaying states.

Before we turn to the effects of the pseudomagnetic field let us inspect some limits.
For vanishing $\gamma_A=\gamma_B=0$ and constant $t_{ab}=t_0$, the system
is periodic and the band structure displays the familiar Dirac cones
$\varepsilon=\pm v|\mathbf{q}|$ near each corner of the Brillouin zone
(the K and K$'$ points situated at $\mathbf{K}_+$ and $\mathbf{K}_-$,
respectively), where $\mathbf{q}=\mathbf{k}- \mathbf{K}_\sigma$ is the
wave vector relative to the corner point \cite{wallace}. Weak uniform
strain, with $t_{ab}=t_{1,2,3}$ only depending on the bond orientation,
displaces the cones from the corners by an amount ${\bf A}$
\cite{Sasaki,saito}. In the presence of amplification and absorption with
$\gamma_A=-\gamma_B=\gamma$, the full band structure of the uniformly
strained system can still be real since in this case the non-hermitian
Hamiltonian \eqref{eq:dirac}
 displays the $\mathcal{PT}$ symmetry
\begin{equation}
\mathcal{H}(x,y)=\sigma_x\mathcal{H}^*(-x,-y)
\sigma_x\equiv \mathcal{PT}H(x,y)\mathcal{PT},
\end{equation}
where $\sigma_x$ is the Pauli matrix. However, when $\gamma$ exceeds a
threshold eigenstates cease to be joint eigenstates of $\mathcal{PT}$,
which leads to complex
branches of the band structure
\cite{ramezani,tachyons}. If amplification and absorption are imbalanced,
all eigenvalues are shifted by $i\bar\gamma$. This includes the case of
`passive' $\mathcal{PT}$ symmetry, where $\bar\gamma=-|\gamma|$ such that
one sublattice is absorbing and the other sublattice is neutral
\cite{experiment1a}. In these more general cases, a relaxed
$\mathcal{PT}$ symmetry can be stated as
\begin{equation}
\mathcal{H}=\mathcal{PT}\mathcal{H}
\mathcal{PT}+2i\bar\gamma.
\label{eq:pt2}
\end{equation}
The spectrum of such a system is constrained to eigenvalues which either
fulfill ${\rm Im}\,\varepsilon_n=\bar\gamma$, or are paired with another
eigenvalue $\varepsilon_{\bar n}=\varepsilon_n-2i{\rm
Im}\,\varepsilon_n+2i\bar\gamma$. However, strain explicitly breaks the
$\mathcal{PT}$ symmetry, as we explore in the following.

\emph{Landau levels.---} We consider a strain configuration which results
in a constant pseudomagnetic field of strength $\beta$. This follows from
a smoothly varying three-fold symmetric configuration with \cite{guinea}
\begin{equation}
t_l=t_0[1-(\beta/2) \boldsymbol{\rho}_l\cdot \mathbf{r}], \quad l=1,2,3,
\label{eq:tfunc}
\end{equation}
which gives rise to a vector potential
$\mathbf{A}=(\sigma\beta/2)(-y\mathbf{i} +x\mathbf{j})$. Microscopically
$\beta$ depends on the sensitivity of the coupling terms on the
nearest-neighbor spacing, as well as on the strain orientation; here we
assume that this parameter is given. Assuming unless otherwise stated
that $\beta>0$ we write the Hamiltonian as
\begin{align}
\mathcal{H}&=\left(\begin{array}{cc} i\gamma_A & v\sqrt{2\beta}\Pi^\dagger \\ v\sqrt{2\beta}\Pi & i\gamma_B \end{array}\right),
\label{eq:heff}
\\
\Pi
&=\frac{1}{\sqrt{2\beta}}(-i\sigma \beta x/2-i\sigma \partial_x+\beta y/2+\partial_y),
\label{eq:pidef}
\end{align}
where $[\Pi,\Pi^\dagger]=1$ coincides with the algebra of harmonic
oscillator annihilation and creation operators. This delivers a spectrum
of Landau levels, with the zeroth level given by
\begin{equation}
\phi_0=\left(\begin{array}{c} \chi_0 \\ 0 \end{array}\right), \quad \varepsilon_0 = i\gamma_A=i\bar\gamma+i\gamma,
\label{eq:zerothll}
\end{equation}
where $\chi_0=(\beta/2\pi)^{1/2}\exp[-\beta(x^2+y^2)/4+\lambda(\sigma x+iy)-\lambda^2/\beta]$
represents the infinitely degenerate set of Landau states
fulfilling $\Pi\chi_0=0$. With $\chi_m=(m!)^{-1/2}(\Pi^\dagger)^m\chi_0$, $m=1,2,3,\ldots$,
the other Landau levels are given by
\begin{align}
\label{eq:higherll}
\phi_n&=\left(\begin{array}{c} \chi_{|n|} \\ \alpha_n \chi_{|n|-1} \end{array}\right),\,\,
\varepsilon_n = i\bar\gamma+\mathop{\rm sgn}(n)\sqrt{\gamma_n^2 -\gamma^2} ,
\\
\alpha_n&=
\mathop{\rm sgn}(n)\sqrt{1 -\frac{\gamma^2}{\gamma_n^2}}
-i\frac{\gamma}{\gamma_n},
\quad \gamma_n = \sqrt{ 2v^2\beta  |n|},
\label{eq:higherllgamma}
\end{align}
for $n=\pm 1,\pm 2,\pm 3,\ldots$. This spectrum is shown in Fig.~\ref{fig2}.

At vanishing $\gamma_{A,B}$ (thus $\bar\gamma=\gamma=0$), these solutions
reduce to the strain-induced Landau levels studied in the graphene
literature \cite{guinea,vozmediano,levy,yan}. For uniform amplification
or absorption ($\gamma=0$), the levels are shifted by $i\bar\gamma$. For
a finite amplification imbalance ($\gamma\neq 0$), the levels with index
$n$ and  $\bar n=-n$ (with $n\neq 0$) coalesce at a threshold value
$|\gamma|=\gamma_n$ and then bifurcate into a pair of levels which
fulfill the spectral constraints stipulated below Eq.~\eqref{eq:pt2}; in
particular, the average imaginary part of these eigenvalues is given by
$i\bar\gamma$. The zeroth Landau level, however, has an imaginary part
which differs from $\bar\gamma$, and is not accompanied by a partner
state (not even in the other valley). This feature rules out the
existence of any $\mathcal{PT}$-like antiunitary operator which would
commute with the Hamiltonian.

The special nature of the zeroth Landau level can be seen as a direct
consequence of the \emph{parity anomaly} \cite{jackiw,semenoff,haldane},
which in the present context is most conveniently identified by
considering the supersymmetric interpretation of Hamiltonians of the form
\eqref{eq:heff} \cite{jackiw}. Depending on whether one eliminates the B
site or A site wave function, the corresponding eigenvalue equation can
be written as
\begin{subequations}
\begin{align}
(\varepsilon-i\gamma_A)\varphi_A&=(\varepsilon-i\gamma_B)^{-1}2\beta v^2\Pi^\dagger\Pi\varphi_A,
\label{eq:susy1}
\\
(\varepsilon-i\gamma_B)\varphi_B&=(\varepsilon-i\gamma_A)^{-1}2\beta v^2\Pi\,\Pi^\dagger\varphi_B,
\label{eq:susy2}
\end{align}\label{eq:susy}%
\end{subequations}
which provides a simple example of supersymmetric partner potentials.
Both equations deliver the same spectrum, except for the zeroth Landau
level, which only occurs in the spectrum of Eq.~\eqref{eq:susy1}.
This state thus breaks the sublattice symmetry---its wavefunction is localized on
the A sublattice, and for $\gamma_A\neq \gamma_B$ this asymmetry is
reflected by a departure from the overall symmetry of the spectrum about
$i\bar\gamma$. This holds for $\beta>0$, as we have assumed so far. For
$\beta<0$, one needs to modify the definitions of $\Pi$ and $\Pi^\dagger$
such that in the Hamiltonian \eqref{eq:heff} they are effectively
interchanged, and the zeroth Landau level is localized on the B
sublattice, with $\varepsilon_0=i\gamma_B$.

Focussing on a single valley (say around the K point, $\sigma=1$), this
anomaly is fully analogous to the parity anomaly in the problem of
massive Dirac electrons in a magnetic field, which possess an extra state
located at energy $E=mc^2$ or $E=-mc^2$ (depending on the sign of the
field) that breaks the symmetry of the spectrum about $E=0$
\cite{haldane}. For electrons on an ordinary honeycomb lattice this
anomaly is canceled in the K$'$ point \cite{semenoff,haldane}, which can
be related to the K point by either using the $\mathcal{P}$ symmetry
(which interchanges the two sublattices) or the $\mathcal{T}$ symmetry
(which inverts the magnetic field).
In the
present photonic setting the $\mathcal{T}$ operation relating the two
valleys in $\mathbf{k}$-space inverts the sign of the amplification
imbalance $\gamma$; furthermore, the $\mathcal{P}$ operation not only
inverts $\gamma$ but also the direction of the vector potential
\eqref{eq:a}---thus, both symmetries are indeed broken. Instead, the
Hamiltonian \eqref{eq:heff} can be mapped from one valley to the other by
the reflection symmetry $x\to-x$, $\sigma\to-\sigma$. Therefore, the
parity anomaly for the zeroth Landau level is replicated identically in
both valleys.

We now turn to the other Landau levels. These are constrained by the
chiral symmetry ${\cal H}(x,-y)^*=-{\cal H}(x,y)$, which results in the
pairing $\varepsilon_{-n}=-\varepsilon_{n}^*$ of eigenvalues before the
bifurcation threshold, $|\gamma|<\gamma_n$.
The question now arises: Why do these levels also obey
the spectral constraints that are usually associated  with $\mathcal{PT}$-symmetric
systems? In particular, before the bifurcation ${\rm
Im}\,\varepsilon_n={\rm Im}\,\varepsilon_{-n}=\bar\gamma$ and the
associated wave function \eqref{eq:higherll} has equal weight on the A
and B sublattices, $|\alpha_{n}|=|\alpha_{-n}|=1$.
After the bifurcation, $|\alpha_n|=1/|\alpha_{-n}|\neq 1$, and the level
with the larger imaginary part has a larger weight on the more amplifying
sublattice, while the other state is predominantly localized on the
opposite sublattice. These properties are all compatible with a existence
of a generalized time-reversal symmetry $\widetilde{\mathcal{PT}}$
applying to these states.

To identify this symmetry we introduce the basis
\begin{equation*}
{|}m,A\rangle\equiv \left(\begin{array}{c}\chi_m \\ 0 \end{array}\right),\quad
{|}m,B\rangle\equiv \left(\begin{array}{c}0 \\ \chi_m \end{array}\right),\quad
m=0,1,2,\ldots
\end{equation*}
of ordinary Landau states localized on the A or B sublattice (we suppress the degeneracy of these levels).
In this basis, the Hamiltonian \eqref{eq:heff} takes the form
$\mathcal{H}=i\gamma_A {|}0,A\rangle\langle 0,A{|}+\widetilde{\mathcal{H}}$,
where
\begin{align*}
&\widetilde{\mathcal{H}}=\sum_{m=0}^\infty \Big(
i\gamma_A {|}m+1,A\rangle\langle m+1,A{|}
+i\gamma_B {|}m,B\rangle\langle m,B{|}
 \\&+ v\sqrt{2\beta(m+1)}({|}m+1,A\rangle\langle m,B{|}
+{|}m,B\rangle\langle m+1,A{|})\Big)
\nonumber
\end{align*}
is the Hamiltonian in the subspace excluding the zeroth Landau level ${|}0,A\rangle$.
Inspecting the properties of
\begin{align}
\widetilde{\mathcal{P}}&=\sum_{m=0}^\infty ({|}m+1,A\rangle\langle m,B{|}
+{|}m,B\rangle\langle m+1,A{|})
\\
\widetilde{\mathcal{T}}&: \Gamma|m,L\rangle \to \Gamma^*|m,L\rangle,\quad L=A,B,
\end{align}
in the original Hilbert space, we find $\widetilde{\mathcal{T}}^2=1$,
$\widetilde{\mathcal{P}}^\dagger =\widetilde{\mathcal{P}}$,
$\widetilde{\mathcal{P}}^2=1-{|}0,A\rangle\langle 0,A{|}$. Thus
$\widetilde{\mathcal{PT}}\equiv\widetilde{\mathcal{P}}\widetilde{\mathcal{T}}|_{n\neq
0}$ is an antiunitary operator in the space of higher Landau levels.
Furthermore, an explicit calculation now delivers the desired relation
$\widetilde{\mathcal{H}} =
\widetilde{\mathcal{PT}}\widetilde{\mathcal{H}}
\widetilde{\mathcal{PT}}+2i\bar\gamma$ in the space of these levels,
which entails the spectral constraints. This symmetry is of dynamical
origin as its construction makes explicit reference to the eigenstates of
the system. The symmetry does not extend to the zeroth Landau level since
$\widetilde{\mathcal{P}}$ is not unitary if this state is included. While
$\widetilde{\mathcal{P}}\widetilde{\mathcal{T}}\mathcal{H} =
\mathcal{H}\widetilde{\mathcal{P}}\widetilde{\mathcal{T}}-2i\bar\gamma
\widetilde{\mathcal{P}}\widetilde{\mathcal{T}}$, the relation
$\widetilde{\mathcal{P}}\widetilde{\mathcal{T}}\mathcal{H}\widetilde{\mathcal{P}}\widetilde{\mathcal{T}}
= \mathcal{H}-2i\bar\gamma -2i\gamma{|}0,A\rangle\langle 0,A{|}$ again
reveals the special spectral status of this level.

\emph{Applications}.--- Our results for the complex spectrum of Landau
levels find their natural applications in the lasing in a two-dimensional
photonic crystal \cite{laserref1,laserref2,laserref3,laserref4}, and in
the beam propagation in  a photonic lattice \cite{peleg,makris,supmat}.

We first consider the onset of lasing, which occurs when the system is
realized in a two-dimensional photonic crystal, with negligible leakage
into the perpendicular direction. The system becomes unstable towards
lasing when the complex frequency of one of the Landau levels acquires a
positive imaginary part. For fixed $\bar\gamma<0$, the system is passive
at $\gamma=0$ (uniform absorption), but as $|\gamma|$ increases the
zeroth Landau level changes its imaginary part, and so do the other
Landau levels beyond their bifurcation thresholds $\gamma_n$. The lasing
threshold $\gamma_L$ now depends on the sign of $\gamma$. If $\gamma>0$,
the lasing threshold is given by $\gamma_L=|\bar\gamma|$ since the zeroth
Landau level is then located on the amplifying sublattice. For
$\gamma<0$, the first level to meet the real axis is associated with the
pair $n=\pm 1$ involving the first Landau level, with lasing threshold
$\gamma_L=\sqrt{2v^2\beta+\bar\gamma^2}$ [see Eq.~\eqref{eq:higherll}].
In the case $\bar\gamma=0$, the lasing threshold either vanishes (for
$\gamma>0$) or is finite (for $\gamma<0$), depending on whether it is set
by the zeroth or first Landau level. These considerations apply to the
principal strain orientation studied here ($\beta>0$). If $\beta$ takes a negative value, the role of the two sublattices is interchanged
and the zeroth Landau level becomes lasing for $\gamma<0$. Similar
asymmetric threshold scenarios arise when one approaches lasing by
changing $\bar\gamma$ at fixed $\gamma$, or when one changes the
amplification rate on one sublattice only. In the lasing regime a
macroscopic number of modes in the zeroth or first Landau level will
participate in the mode competition. In a finite system, the exact
degeneracy will be lifted by the boundary conditions, but this lifting
will be small in the bulk, while edge states can also appear; disorder in
the couplings and amplification rates will also broaden the levels.

When the system is realized in an array of single-mode waveguides the
propagation along these waveguides is free, and the eigenvalues represent
complex propagation constants, where the imaginary part describes the
spatial decay or increase of the eigenmodes \cite{peleg,makris,supmat}.
An attractive feature of such settings is the possibility to probe the
parity anomaly in a system without any active (amplifying) components.
For this we set, e.g., $\gamma_A <0$ and $\gamma_B=0$, such that one
sublattice is lossy while the other sublattice is passive. The parity
anomaly can then be probed via a beam fed into one end of the waveguide
array. The output beam will show the component of the zeroth Landau level
either unaffected, or suppressed according to a  decay constant
$|\gamma_A|$. Assuming that $|\gamma_A|/2<\gamma_1$, all other components
are suppressed uniformly according to a common decay constant
$|\gamma_A|/2$. The initial population of the modes can be controlled via
the input beam. This approach mirrors passive implementations of
$\mathcal{PT}$-symmetric optics and avoids complications from the
intrinsic dispersion in the active parts
\cite{experiment1a,experiment1b}.

\emph{Conclusions.}---In summary, we described the formation of Landau
levels in a strained photonic system and identified means to probe the
associated  parity anomaly via  amplification that breaks the sublattice
symmetry. We focussed the attention on the honeycomb lattice as it
naturally provides a conical dispersion, a pseudomagnetic field under
strain, and two sublattices which can be equipped with gain and loss. It
should be pointed out that conical dispersions are a generic feature of
triangular lattices with inversion and time-reversal symmetry
\cite{raghu}, while the correspondence between strain and a
pseudomagnetic field generalizes to other lattices \cite{iordanskii}.
Guided by coupled-mode theory \cite{peleg,makris,supmat} one can identify
a number of lattices with Dirac-like dispersions. Particularly promising
variants include the Lieb lattice (a tripartite lattice which in addition
exhibits a flat band) \cite{lieb} and the one-dimensional
Su-Schrieffer-Heeger chain \cite{ssh,hs}, a  bipartite system which provides
the platform for the recently reported $\mathcal{PT}$-symmetric Talbot
effect \cite{talbot}.

\emph{Note added}.---We end by pointing the reader to recent
experimental work on a passive photonic honeycomb
lattice \cite{Rechtsman}, which has shown that the three-fold symmetric strain pattern is
indeed feasible and results in the formation of the Landau-levels
described here.

\newcommand{\brho}{\boldsymbol{\rho}}
\newcommand{\bk}{\mathbf{k}}

\renewcommand{\theequation}{S\arabic{equation}}
\renewcommand{\thefigure}{S\arabic{figure}}

\appendix

\section*{Supplemental material:
Parity anomaly and Landau-level lasing\\ in strained photonic honeycomb lattices}

The argumentation in the main text is based on a Dirac-like wave equation
(1) which incorporates gain, loss and nonuniform strain. Here we
describe how this equation emerges from a microscopic model of a
photonic honeycomb structure. We first focus on the technical details,
which amount to a synthesis of works on graphene
\cite{wallace,Sasaki,saito,guinea,vozmediano} and $\mathcal{PT}$-symmetric
lattices
\cite{peleg,makris,experiment1a,experiment1b,tachyons,ramezani,Regensburger,talbot,Rechtsman},
and then discuss the interpretation of the result.

\begin{figure}
\includegraphics[width=\columnwidth]{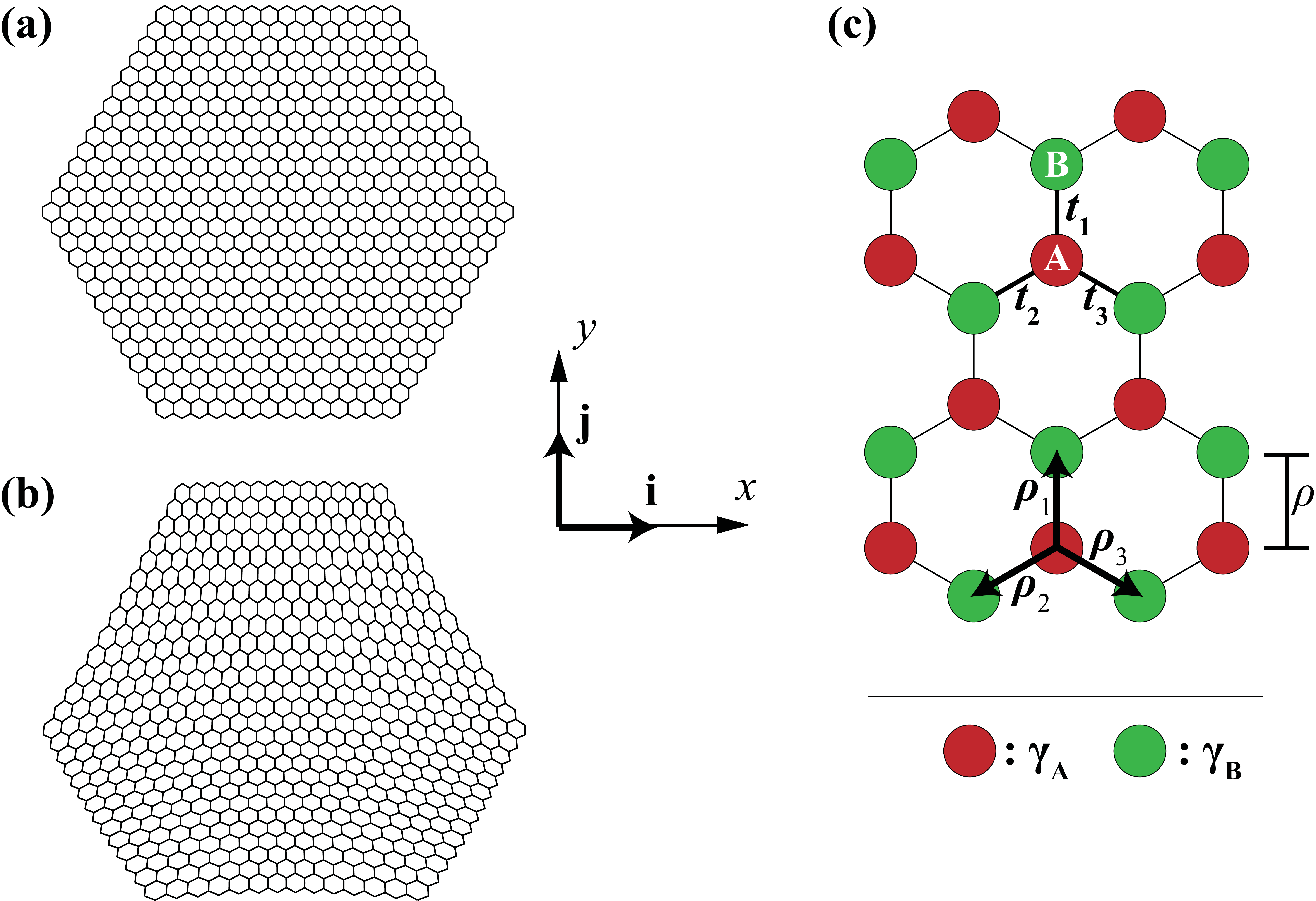}
\caption{\label{figs1} Replication of Fig. 1 in the main text.
(a) Segment of a honeycomb lattice,  with vertices representing states in a two-dimensional photonic crystal or weakly coupled optical fibers in a three-dimensional setting.
(b) Sketch of a deformed arrangement which results in a constant pseudomagnetic field.
(c)
The two sublattices A and B have amplification rates $\gamma_A$ and $\gamma_B$,
respectively (negative values correspond to absorption).
The pseudomagnetic field resulting from the strain is modeled
via smooth coupling functions $t_l$ for bonds aligned along the vectors $\boldsymbol{\rho}_l$, $l=1,2,3$.}
\end{figure}

As in these previous investigations
we base the microscopic considerations on coupled-mode theory. In this
theory, a tight-binding model is formulated on a lattice, where each
vertex is associated with a localized mode while the bonds represent the
coupling between the modes. This is illustrated in Fig.~\ref{figs1}, which
replicates Fig.~\ref{fig1} in the main text. Panel (a) shows a segment of a
honeycomb lattice, consisting of two sublattices A and B. We denote the
associated modes by $|a\rangle$, and $|b\rangle$, where the indices $a$
run over the A sublattice and the indices $b$ run over the B sublattice.
Panel (b) sketches the deformed arrangement which results in a constant
pseudomagnetic field. Panel (c) explains the microscopic modeling of
these effects, which we elaborate in these supplemental notes. The modes
on each sublattice are equipped with amplification or absorption rates
$\gamma_A=\bar\gamma+\gamma$ and $\gamma_B=\bar\gamma-\gamma$. Here
$\bar\gamma$ is the average rate and $\gamma$ quantifies the imbalance.
We only consider nearest-neighbor couplings and denote the coupling rates
as $t_{ab}$. In the unstrained case, $t_{ab}=t_0$ is constant. Strain
results in a spatial variation of the coupling terms, which we
parameterize as $t_{ab}=t_{l}(\mathbf{r}_a)$, where $\mathbf{r}_a$ is the
unstrained position of the A site and $l=1,2,3$ indicates the orientation
along the unstrained bond vectors $\brho_l$ (with $|\brho_l|=\rho$ the
unstrained nearest-neighbor distance). We orientate the lattice such that
the orthogonal cartesian unit vectors
$\mathbf{i}=(\brho_3-\brho_2)/\sqrt{3}\rho$ and
$\mathbf{j}=\brho_1/\rho$.

\begin{figure}
\includegraphics[width=.6\columnwidth]{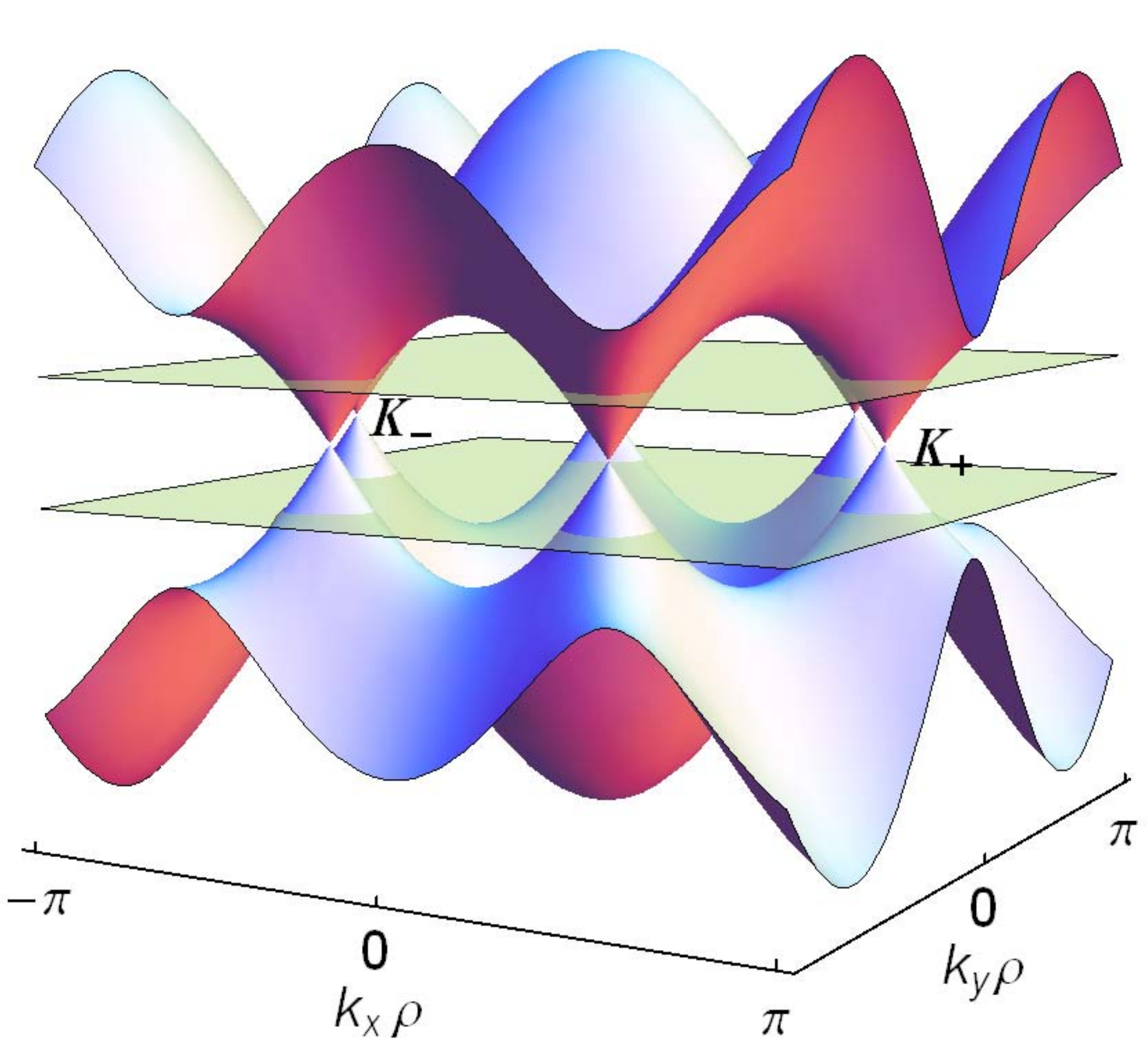}
\caption{\label{figs2} Dispersion of the unstrained passive honeycomb lattice in coupled-mode theory. The planes focus onto the central region, formed by  cones with apex at the corners of the Brillouin zone (represented by the $\mathbf{K}_+$ and $\mathbf{K}_-$ points).}
\end{figure}

Based on the quantities introduced above, coupled-mode theory determines
the eigenmodes $|\psi\rangle=\sum_a\psi_a|a\rangle+\sum_b\psi_b|b\rangle$
of the system as the eigenstates of an effective Hamiltonian
\begin{equation}
\label{eq:tb}
H=i\gamma_A\sum_{a} {|}a\rangle\langle a{|} +i\gamma_B\sum_{b} {|}b\rangle\langle b{|}-\sum_{{<}ab{>}}\!\!t_{ab}({|}a\rangle\langle b{|}+{|}b\rangle\langle a{|})
\end{equation}
(where the negative sign in front of the coupling terms follows convention).
The eigenvalue equation is $\varepsilon|\psi\rangle=H|\psi\rangle$; the interpretation of the eigenvalues $\varepsilon$ depends on the context and is discussed towards the end of these notes.

We are interested in the modes of a system with smoothly varying coupling constants, in a spectral range close to the center of the dispersion of the unstrained passive system.
The Dirac equation used in the text then arises in a gradient expansion in the lattice indices.  To obtain the reference point for the expansion we set $t_{ab}=t_0$ and $\gamma_A=\gamma_B=0$. The eigenmodes are then of Bloch form
\begin{eqnarray}
|\psi\rangle&=&\sum_a\phi^{(A)}\exp[i{\bf k}\cdot ({\bf r}_a+\brho_1/2)]|a\rangle
\nonumber \\
&&+
\sum_b\phi^{(B)}\exp[i{\bf k}\cdot ({\bf r}_b-\brho_1/2)]|b\rangle,
\end{eqnarray}
where we follow the convention to reference the lattice points to a suitably chosen center of a unit cell encompassing an A and a B site at the ends of a vertical bond. Inserting this ansatz into the eigenvalue equation one arrives at the Bloch Hamiltonian
\begin{equation}
\label{blochham}
\mathcal{H}=\left(
    \begin{array}{cc}
      0 & -f \\
      -f^* & 0 \\
    \end{array}
  \right),
\end{equation}
where
\begin{equation}
f=t_0[1+e^{i\mathbf{k}\cdot(\brho_2-\brho_1)}+e^{i\mathbf{k}\cdot(\brho_3-\brho_1)}].
\end{equation}
The associated dispersion relation $\varepsilon=\pm|f({\bf k})|$, plotted in Fig.~S\ref{figs2},
is conical in the center of the band, with the apex of each cone situated at a corner of the hexagonal Brillouin zone
\cite{wallace}.
Only two of these $K$ points are independent, while the others are related by reciprocal lattice vectors. We set  $\mathbf{K}_\sigma=\sigma (4\pi/3\sqrt{3}\rho)\mathbf{i}$ and distinguish the two
independent choices by the valley index $\sigma=\pm1$.
Expanding the wave vector around these points, ${\bf k}= {\bf K}_{\sigma}+{\bf q}$ with ${\bf q}=q_x{\bf i}+q_y{\bf j}$, one
obtains
\begin{eqnarray}
f&\approx&
-\sigma v q_x
+ i v q_y
\end{eqnarray}
where $v=\frac{3 t_0\rho}{2}$.
This delivers the conical dispersion $\varepsilon=v|{\bf q}|$ about the K points.

We now incorporate gain, loss and non-uniform but smooth strain within a gradient expansion, which captures these effects via a smooth envelope function that modulates the Bloch wave function.
The full spatial dependence of the wave function is of the form
\begin{eqnarray}
\label{eq:gradient}
|\psi\rangle&=&\sum_a\phi^{(A)}({\bf r}_a)\exp[i{\bf K}_{\sigma}\cdot ({\bf r}_a+\brho_1/2)]|a\rangle
\nonumber\\
&&+
\sum_b\phi^{(B)}({\bf r}_b)\exp[i{\bf K}_{\sigma}\cdot ({\bf r}_b-\brho_1/2)]|b\rangle
\end{eqnarray}
where we separated the rapid oscillations with wave number ${\bf K}_{\sigma}$ from
the slowly varying envelope functions $\phi^{(A)}({\bf r})$ and $\phi^{(B)}({\bf r})$. For the unstrained passive system,
comparison with the expressions above delivers
\begin{eqnarray}
\phi^{(A)}({\bf r}_a)&=\phi^{(A)}\exp[i{\bf q}\cdot ({\bf r}_a+\brho_1/2)],
\nonumber\\
\phi^{(B)}({\bf r}_b)&=\phi^{(B)}\exp[i{\bf q}\cdot ({\bf r}_b-\brho_1/2)],
\end{eqnarray}
where $(\phi^{(A)},\phi^{(A)})^T$ is an eigenvector of the Bloch Hamiltonian
\eqref{blochham}.
The gradient expansion adapts
these functions to the case of couplings $t_{ab}=t_l(\mathbf{r}_a)$
with functions $t_l$, $l=1,2,3$, that vary smoothly across the lattice ($|\nabla t_l| \rho\ll t_0$).
If these functions were constant we would arrive at the Bloch Hamiltonian \eqref{blochham} with
\begin{eqnarray}
f&=&t_1+t_2e^{i\mathbf{k}\cdot(\brho_2-\brho_1)}+t_3e^{i\mathbf{k}\cdot(\brho_3-\brho_1)}
\nonumber\\
&\approx&
-\sigma v (q_x-A_x)
+i v  (q_y-A_y)
,
\label{eq:fa}
\end{eqnarray}
where we again expanded about a K point and abbreviated
\begin{equation}
A_x=\sigma \frac{1}{3t_0\rho}(2t_1-t_2-t_3),\quad
A_y=\sigma \frac{1}{\sqrt{3}t_0\rho}(t_2-t_3).
\end{equation}
To capture the variations
we insert Eq.\ \eqref{eq:gradient} into the eigenvalue equation of the microscopic coupled-mode Hamiltonian
\eqref{eq:tb}
and determine the amplitudes of the neighboring unit cells via the Taylor expansion
$\phi^{(A,B)}({\bf r}+\Delta {\bf r})\approx [1+\Delta {\bf r}\cdot\nabla]
\phi^{(A,B)}({\bf r})$, where $\Delta \mathbf{r}$ is the lattice vector connecting the centers of the
unit cells in question. Locally, the Hamiltonian is then still of the form \eqref{blochham} with $f$ given as in \eqref{eq:fa},
but with $q_x\to p_x\equiv -i\partial_x$ and $q_y\to p_y\equiv -i\partial_y$.
In the last step of the derivation we account for the amplification and absorption rates $\gamma_A$ and $\gamma_B$. These are constant throughout the lattice, thus directly lift from the coupled mode equations to the Dirac Hamiltonian,
which takes the final form
\begin{equation}
\mathcal{H}=\left(\begin{array}{cc} i\gamma_A & v (\sigma P_x-iP_y)\\ v (\sigma P_x+iP_y) & i\gamma_B \end{array}\right),
\end{equation}
$P_x=-i\partial_x-A_x$, $P_y=-i\partial_y-A_y$;
see Eq.~(1) of the main text.

\begin{figure}
\includegraphics[width=\columnwidth]{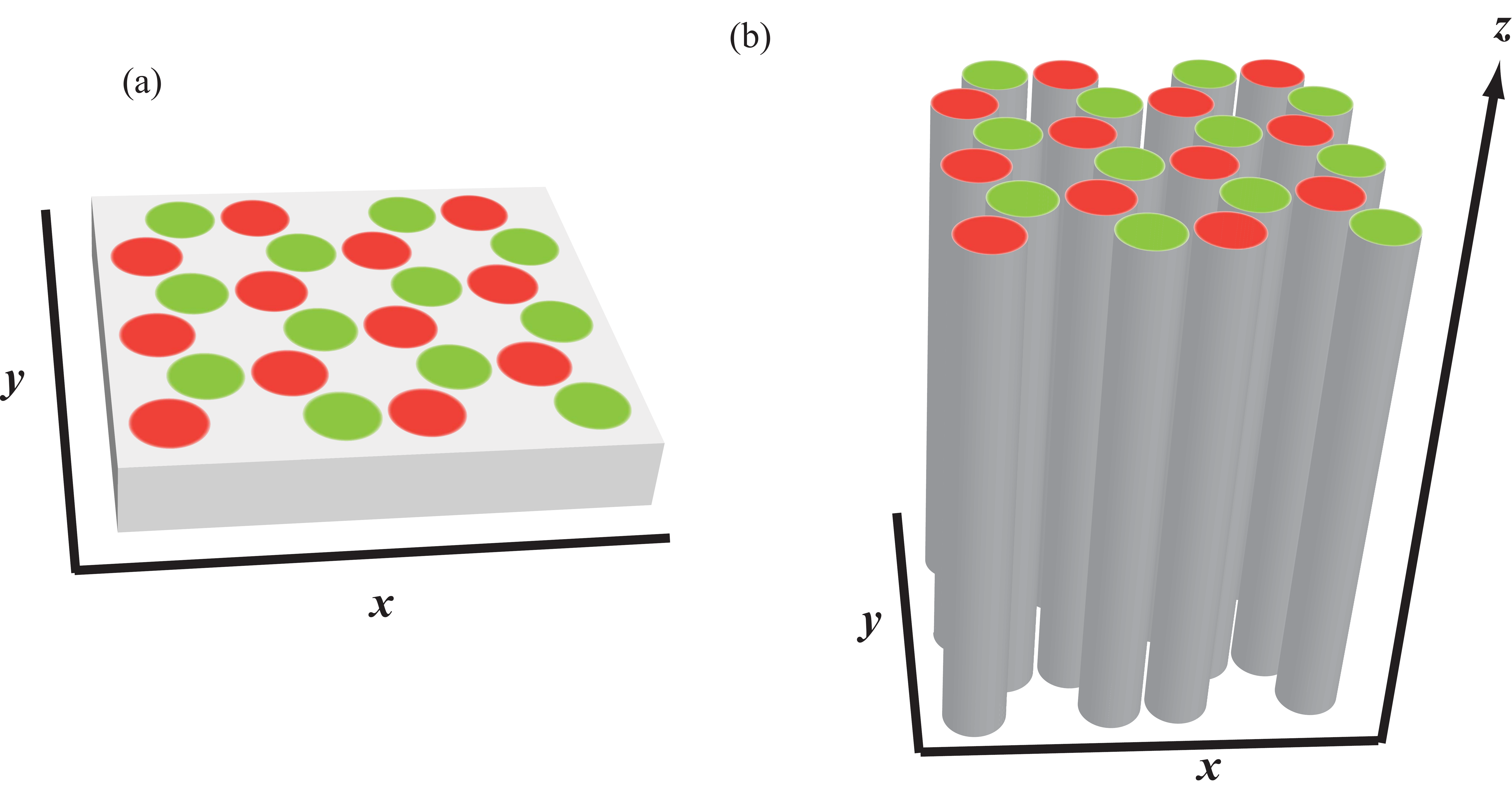}
\caption{\label{figs3} Sketch of photonic systems with underlying honeycomb structure: (a) Two-dimensional photonic crystal, (b) Photonic lattice.}
\end{figure}

In the main text, we applied this description to two different physical
settings, a two-dimensional photonic crystal as in Fig.~\ref{figs3}(a), or a
photonic lattice of  single-mode waveguides as in Fig.~\ref{figs3}(b). Depending
on the setting the modes $|a\rangle$, $|b\rangle$ can then be associated,
e.g., to Wannier states which support the modes in a two-band
approximation, or the wave-guide modes which are weakly coupled via
tunneling. For the two-dimensional crystal  the Hamiltonian is then
interpreted as the generator of the time evolution in $t$, while for the
 photonic lattice it generates the propagation into the $z$ direction
\cite{peleg,makris}. Accordingly, the eigenvalues $\varepsilon$ of $H$
determine the frequencies $\omega$ of quasibound states in the two
dimensional crystal or the propagation constant $ck_z$ along the third
direction in the photonic waveguide lattice. Eigenvalues with a positive
imaginary part correspond to amplified states  (in time or along the
propagation direction, respectively), while those with a negative
imaginary part correspond to decaying states. This interpretation can be
made explicit when one focusses on a small window of the eigenvalue
spectrum centered around a large value $\varepsilon_0$, which induces a
rapid modulation $\exp(-is\varepsilon_0)$ with $s=z$ or $t$. The
underlying second-order wave equation (e.g., the Helmholtz equation or
Maxwell's equations) can then be simplified in a gradient expansion,
which amounts to the substitution $-\partial_s^2\to
(\varepsilon_0+i\partial_s)^2\approx
\varepsilon_0^2-2i\varepsilon_0\partial_s$. The first term is an offset
which centers the spectrum around $\varepsilon_0$, so that the evolution
equation for the $s$-dependent envelope function reads
$i\partial_s\psi(x,y,s)=H\psi(x,y,s)$. For $s=z$, this is the paraxial
approximation; and it is indeed this interpretation which has guided
theory and experiment on $\mathcal{PT}$-symmetric
\cite{peleg,tachyons,makris,experiment1a,experiment1b,ramezani,Regensburger,talbot}
and strained \cite{Rechtsman} photonic lattices.

\end{document}